 \documentclass[final,5p,times,twocolumn,authoryear]{elsarticle}

\usepackage{amssymb}
\usepackage{lipsum}
\usepackage{arydshln}
\usepackage{lscape}
\usepackage{float}

\journal{Natural Language Processing Journal}

\begin{document}

\begin{frontmatter}



\title{Evaluating LLMs on Document-Based QA: Exact Answer Selection and Numerical Extraction using CogTale dataset}

\author[affil1]{Zafaryab Rasool\corref{cor1}}
\ead{zafaryab.rasool@deakin.edu.au}

\author[affil1]{Stefanus Kurniawan }
\ead{stefanus.kurniawan@deakin.edu.au}

\author[affil1]{Sherwin Balugo }
\ead{s.balugo@deakin.edu.au}

\author[affil1]{Scott Barnett }
\ead{scott.barnett@deakin.edu.au}

\author[affil1]{Rajesh Vasa }
\ead{rajesh.vasa@deakin.edu.au}

\author[affil2]{Courtney Chesser }
\ead{courtney.chesser@deakin.edu.au}

\author[affil3]{Benjamin M. Hampstead }
\ead{bhampste@med.umich.edu}

\author[affil4]{Sylvie Belleville }
\ead{sylvie.belleville@umontreal.ca}

\author[affil1]{Kon Mouzakis }
\ead{kon.mouzakis@deakin.edu.au}

\author[affil2]{Alex Bahar-Fuchs }
\ead{a.baharfuchs@deakin.edu.au}


\address[affil1]{Applied Artificial Intelligence Institute, Deakin University, Melbourne, Australia}
\address[affil2]{School of Psychology, Faculty of Health, Deakin University, Applied Artificial Intelligence Institute, Deakin University, Melbourne, Australia}
\address[affil3]{Michigan Alzheimer's Disease Research Center, University of Michigan, Ann Arbor, USA}
\address[affil4]{Psychology Department, Université de Montréal, Quebec, Canada}

\cortext[cor1]{Corresponding author}

\begin{abstract}
Document-based Question-Answering (QA) tasks are crucial for precise information retrieval. While some existing work focus on evaluating large language model’s performance on retrieving and answering questions from documents, assessing the LLMs’ performance on QA types that require exact answer selection from predefined options and numerical extraction is yet to be fully assessed. In this paper, we specifically focus on this underexplored context and conduct empirical analysis of LLMs (GPT-4 and GPT-3.5) on question types, including single-choice, yes-no, multiple-choice, and number extraction questions from documents in zero-shot setting. We use the CogTale dataset for evaluation, which provide human expert-tagged responses, offering a robust benchmark for precision and factual grounding. We found that LLMs, particularly GPT-4, can precisely answer many single-choice and yes-no questions given relevant context, demonstrating their efficacy in information retrieval tasks. However, their performance diminishes when confronted with multiple-choice and number extraction formats, lowering the overall performance of the model on this task, indicating that these models may not yet be sufficiently reliable for the task. This limits the applications of LLMs on applications demanding precise information extraction from documents, such as meta-analysis tasks. These findings hinge on the assumption that the retrievers furnish pertinent context necessary for accurate responses, emphasizing the need for further research. Our work offers a framework for ongoing dataset evaluation, ensuring that LLM applications for information retrieval and document analysis continue to meet evolving standards.
\end{abstract}



\begin{keyword}
Large Language Models \sep Document-based information retrieval \sep Evaluation \sep Question-Answering \sep CogTale dataset \sep healthcare



\end{keyword}

\end{frontmatter}

\section{Introduction}
Large language models (LLMs) have recently gained attention due to their ability to solve various natural language processing tasks (\cite{espejel2023gpt, aher2023using, acharya2023llm, zhao2023survey}). However,  existing evaluation of LLMs predominantly focuses on general knowledge questions and reasoning tasks \cite{bian2023chatgpt, qin2023chatgpt, bai2023benchmarking, bang2023multitask}, rather than retrieval of specific information from documents. 
In real-world, various scenarios require extracting the number of participants in the control group of a trial in a medical paper, relevant policy information from policy documents, specific dollar value of liability in a legal context, and so on. The current favoured approach to solve these tasks involve using Retrieval Augmented Generation (RAG). However, the effectiveness of LLMs on these narrow tasks is under explored, which limits the evaluation of these system's real-world applicability. Furthermore, while existing datasets tend to focus on the performance of LLM, document QA datasets around RAG are in their infancy.



\begin{table*}[!t]
    \centering
    \begin{tabular}{|p{18cm}|}
    \hline

             Question: \textit{Was the intervention delivered as per the planned protocol? i.e., no significant changes to  the protocol implemented after the trial began?} \\
         Category: Yes-No type\\
         Options: [Yes, No, Not specified]\\
         Actual answer:  \\
         \\
         \hdashline
Question: \textit{What type of trial was conducted to evaluate the intervention?} \\
Category: Single-choice\\
Options: [ Randomised controlled trial- parallel groups  ,  Randomised controlled trial- cross over trial  ,  Randomised controlled trial- cluster ,  Randomised controlled trial -Waitlist-control ,  Non randomised controlled trial ,  Open (before and after) trial (no control) ,  Single case (with phase randomization) ,  Single case (without phase randomization) ,  Randomized interventional study (no control group) ,  Partial-randomized controlled trial ,  Parallel groups]\\
Actual answer: \\
    \\
    \hdashline
    
         Question: \textit{What is the number of control conditions?}\\
         Category: Single-choice (number)\\
         Option: [0, 1, 2, ..., 21]\\
        Actual answer: \\
        \\
        \hdashline
        
         Question: \textit{Which individuals were deliberately kept unaware of the specific intervention they received in the study?}\\
         Category: Multiple-choice\\
         Option: [Assessors, Trainers/therapists, Participants, Data analysts, No blinding attempted,  Not specified, N/A, Caregivers]\\
         Actual answer: \\
\\
         \hdashline
         Question: \textit{What proportion of participants from the control group were retained at the post-intervention assessment?}\\
         Category: Number-extraction\\
         Actual answer:  \\

        \hline

    \end{tabular}
    \caption{Example questions from the CogTale dataset belonging to different question-type category}
    \label{tab:cogtale}
\end{table*}


We take the example of the CogTale platform \cite{sabates2021cogtale} as a running example which consist of database of published research papers on cognitive interventions for older adults. Researchers interested in evaluating the quality of trials entered onto the CogTale database and synthesizing the evidence generally perform manual annotation and data extraction into the database using a structured form . However, the manual retrieval/extraction of target information from these documents is a laborious process, potentially leading to challenges such as mis-interpretation, scalability issues for projects with stringent timelines, inconsistencies, and the potential for errors, which slows down the evidence translation and implementation process.  LLMs which have proved their effectiveness on several tasks such as summarizing, reasoning and others, can offer potential solution to the aforementioned issues. Therefore, there is a need to investigate the performance of LLMs in information retrieval tasks.


 Existing related work by \cite{pereira2023visconde} evaluated GPT-3's performance on the above task
 using three datasets (IIRC, Qasper and StrategyQA). These dataset mostly focus on complex context comprehension and multi-paragraph answer extraction.  Other popular datasets such as PubMedQA (\cite{jin-etal-2019-pubmedqa}) and BioASQ (\cite{krithara2023bioasq}) involve either asking yes-no type question or factoid and list questions. 
 However, how LLMs perform on question types that require selecting answers from provided response options and extracting numerical values is not yet fully explored. Such question-answer formats are prevalent in various scenarios, including in healthcare-related evidence synthesis tasks, and gaining insights into LLMs performance in these areas would enable users to confidently employ them for such tasks. Examples of such questions are shown in Table \ref{tab:cogtale}.

 
Therefore, in this paper, we focus on the task of extracting specific information from the CogTale dataset using LLM, specifically GPT-3.5-turbo and GPT-4 (\cite{openai2023gpt4}). We developed a pipeline which involves extracting related passages from document(s) based on the question, and prompting an LLM to select the correct answer(s) from a set of options using the extracted passages. CogTale data extraction form  consists of a set of questions that can be categorised into single choice, multiple choice, single choice (number options), yes-no type, etc. Additionally, direct value/number extraction and value computation questions are also included. These questions along with related passages extracted from the document(s) are passed to an LLM for generating the answers.

We conduct an empirical analysis on 13 studies, consisting of the research papers and the different question types selected from the Cogtale platform, using the above developed pipeline in a zero-shot setting. Based on the analysis, we found that GPT-4 surpassed GPT-3.5-turbo in performance across all question types from CogTale dataset. However, the overall performance of these models was not found satisfactory as GPT-4 achieved an overall accuracy of 41.84\%, which shows that these models may not be reliable. In terms of the different categories of questions, GPT-4 performed better on single-choice questions and yes-no type questions as compared to multiple-choice and number-extraction. This demonstrates that the current versions of GPT may not be reliable for these tasks in a zero-shot setting, under the assumption that retrievers provided essential context for the task.

Based on the above, we summarise our contributions in this paper as follows:

\begin{itemize}

\begin{table*}[!t]
    \centering
    \begin{tabular}{|l|c|c|c|c|c|}
    \hline
       Dataset  & Single-choice & Multiple-choice & Single-choice numbers & Yes-No & Number extraction \\
       \hline
        IIRC & - & - & - & - & -\\
        StrategyQA & - & - & - & Yes & -\\
        Qasper & - & - & - & Yes & -\\
        PubMedQA & - & - & - & Yes & - \\
        BioASQ & - & - & - & Yes & - \\
        CogTale & Yes & Yes & Yes & Yes & Yes \\
        \hline
    \end{tabular}
    \caption{Comparison of different QA datasets based on their question-types. Here \lq -' indicates that the particular category type is not present or not covered specifically in the dataset.}
    \label{tab:compare_dataset}
\end{table*}

\item \textbf{Diverse Question formats:} We conducted experimental analysis of Large Language Models (LLMs) with a focus on GPT-4 and GPT-3.5-turbo across diverse question formats such as yes-no, single-choice, multiple-choice, number-extractions, in the context of document-based information retrieval.

\item \textbf{Utilization of CogTale dataset:} Leveraged the CogTale dataset, featuring research papers on cognitive interventions for older adults, to demonstrate the practical applicability of LLMs in retrieving information from documents, offering valuable insights for researchers and practitioners in healthcare and related fields.
\end{itemize}

In the remainder of the paper, we first discuss the background in Section 2 and then present the Methodology in Section 3 covering the details of the dataset and the QA framework. Empirical evaluation and analysis are discussed in Section 4. We provide a discussion of the results in Section 5. Finally, we present the conclusion and future work in Section 5, and threat to validity in Section 7.

\section{Background}






A notable surge in research endeavors has been directed towards the exploration of large language models recently. This surge has seen numerous studies evaluating LLMs performance on different tasks as highlighted in recent surveys  (\cite{zhao2023survey}, \cite{chang2023survey}, \cite{kalyan2023survey}). Most existing works evaluate the performance of LLMs on benchmark and open-domain questions focused on reasoning and factoid questions.  \cite{bian2023chatgpt} evaluated the performance of ChatGPT on commonsense problems from different domains. \cite{qin2023chatgpt} investigated the zero-shot performance of ChatGPT and GPT 3.5 on several NLP tasks. \cite{bai2023benchmarking} propose to use the language model as a knowledgeable examiner which evaluates other models on the responses to its questions. \cite{bang2023multitask} evaluates ChatGPT on NLP tasks. \cite{kamalloo2023evaluating} evaluates LLMs and other open-domain QA models by manually evaluating their answers on a benchmark dataset. 

Information retrieval using LLMs have gained attention recently (\cite{ram2023context, shi2023replug, levine2022huge}). A recent work by \cite{pereira2023visconde} evaluates GPT-3's performance on three information-seeking datasets including the Incomplete Information Reading Comprehension (IIRC) Questions dataset (\cite{ferguson2020iirc}), QASPER dataset (\cite{dasigi2021dataset}) and StrategyQA dataset (\cite{geva2021did}). \cite{liu2023gradually} also evaluated their approach on StrategyQA dataset. While the IIRC and StrategyQA dataset are focused on complex context comprehension and multi-paragraph evidence extraction, QASPER dataset consist of research papers focused on natural language processing topics and question types such as: Extractive, Abstractive, Yes/No and Unanswerable.


CogTale dataset differs from the above datasets and other popular datasets such as PubMedQA and BioASQ. PubMedQA (\cite{jin-etal-2019-pubmedqa}) focuses on biomedical research papers and uses the abstract of their research papers for questions, while CogTale uses data extracted/retrieved from complete papers. BioASQ (\cite{paliouras2014challenge}) focus on open-domain QA over PubMed abstracts and include yes-no type, factoid and list questions. Cogtale specifically focuses on question-types such as selecting single or multiple correct answer(s) from a list of options specially, which differentiates it from other existing datasets (Table \ref{tab:compare_dataset} provides a summary of the comparison of the above datasets and the Cogtale dataset). While the datasets may involve some questions of these types, they are not specifically focused on such types. Thus, bridging this gap by evaluating LLM’s performance on CogTale dataset is crucial as it addresses a fundamental aspect of information retrieval, enabling context-aware and efficient access to knowledge from documents.

\section{Methodology}

We first describe the details of the CogTale dataset and then discuss the framework to evaluate the performance of LLMs on question-answering tasks. 





      

\subsection{Details of CogTale Dataset}



The CogTale platform is a repository of methodological and outcome data from trials of cognition-oriented treatments for the elderly and was developed as part of efforts to semi-automate key aspects of the evidence synthesis pipeline. CogTale serves as a valuable resource for researchers and clinicians seeking related information about trials and/or interested in rapid evidence synthesis. Facilitated by the CogTale platform, users can seamlessly search for specific studies and access precise details from them. Furthermore, the platform enables users to contribute their own studies, establishing an efficient medium for information retrieval. The platform include a wealth of data for each study included in the dataset, such as trial specifications, total sample size and its rationale, eligibility criteria, primary and secondary outcomes, intervention particulars, study findings, and more. Next, we discuss the question-types.






\medskip
\noindent
\textbf{Question types:} The CogTale data extraction form  comprises a diverse array of questions, organized into eight different sections, focused on very specific information from different studies (i.e., research papers). The same question set is used to extract information from all studies in the database and seek information about how a trial was conducted, number of participants, and other useful information. We classify these questions into distinct types, providing detailed elaboration below.


\begin{enumerate}
\item  \textbf{Yes-No type}: This question category requires a response in the form of \lq yes" or \lq no." This category may also include options such as \lq Not Specified", \lq N/A" or other similar options.

\item  \textbf{Single-choice}: The second type involves questions accompanied by multiple options, with a singular correct answer among them.

\item  \textbf{Single-choice (number)}: This category includes options that involve numerical value as answer, and one of the answer is correct.

\item  \textbf{Multiple-choice}: In this category, a question consists of many options and more than one options can be correct.

\item  \textbf{Number-extraction}: In this category, the expected response is a numerical value. However, this category does not involve options which distinguishes it from the third category: single-select from options (number).

\end{enumerate}

Examples of these question type from the CogTale dataset are shown in Table \ref{tab:cogtale}. 


\subsection{Question-Answering (QA) Framework}
We explain the QA framework using document-based QA dataset (particularly CogTale dataset). The task of retrieving specific information from CogTale dataset can be described as the below problem definition. \textit{Given a question \textit{q} with a list of options and a document \textit{d}, use the LLM to select the answer to question q utilising information from \textit{d} as supporting context.}

\begin{figure}[!t]
\centering
  \begin{center}
    \includegraphics[width=9.5cm,height=6cm]{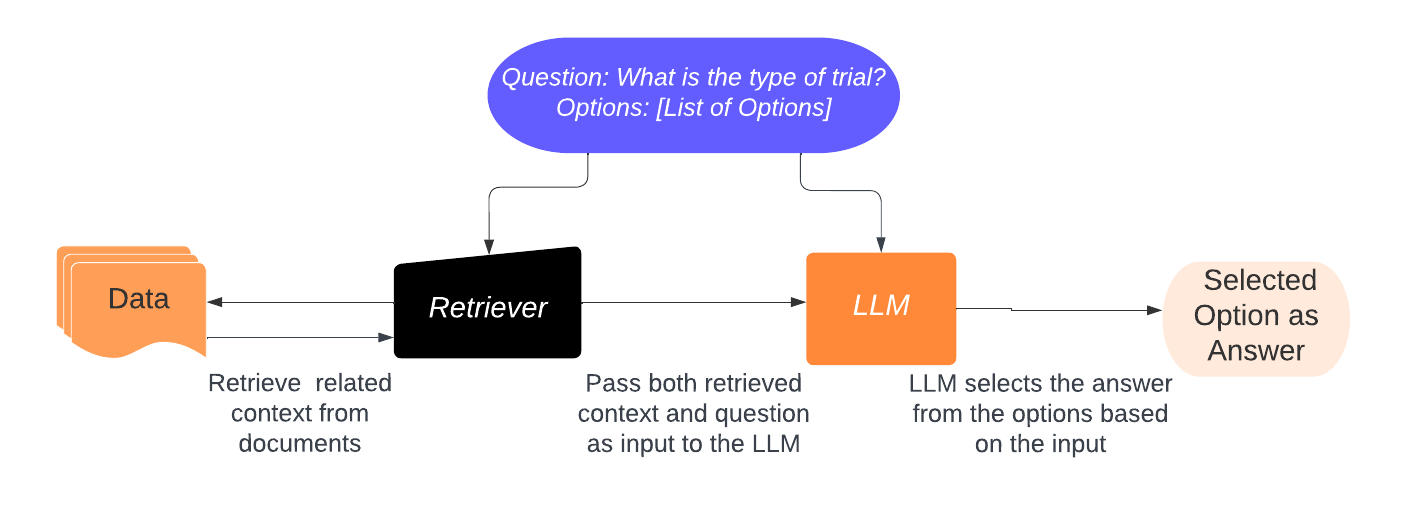}

    \caption{Question-Answering (QA) Framework using LLM for a document-based QA task.}\label{fig:rag}
  \end{center}  
\end{figure}

Based on the above problem definition, the QA framework is illustrated as shown in Figure \ref{fig:rag}. 
Broadly there are two steps in this framework: (1) Retrieve, and (2) Answer. In the first step, a retriever is responsible for retrieving the most relevant information from the document based on the input question. This is an important step as it ensures that the model considers the appropriate context when generating answers. 
Initially, the document is divided into chunks, and then embeddings are generated for each chunk using an embedding model. When a new question comes, the retriever uses the question embeddings to find the relevant document chunks. This is a similarity task as most similar chunks to the question are required. Thus, an appropriate similarity measure (such as Cosine Similarity) is used to extract the most relevant chunks. 

The selected chunks, along with the question, are passed to an LLM (GPT-3.5-turbo and GPT-4) to generate the answer. Since most questions in our study focus on selecting answer(s) from options, the LLM is required to select one or multiple correct answer based on the question-type. For this task, an LLM needs to be prompted and appropriate prompting is essential. For this study, we utilize straightforward prompts that explicitly outline the question's requirements. Our prompts to the model consist of a question presented alongside answer options, as well as the corresponding passages extracted from the document. For the different question types we discussed earlier, the example of prompts are shown in the tables \ref{tab:my_label1}, \ref{tab:my_label2}  and \ref{tab:my_label3}.

 \begin{table*}[!t]
    \centering
    \begin{tabular}{|p{18cm}|}
\hline
          Use the following pieces of context to extract the information at the end. If you can't find the answer, just say that you don't know, don't try to make up an answer.\\   
\{summaries\}  \\
\\  
You can only answer from one of these values:\\

\{answer\_options\}  \\
\\
 Question: \{question\}  \\
\\
The answer can only be the exact value of one of the options. Just return the final value when answering.\\

Answer:\\
\hline 

    \end{tabular}
    \caption{Prompt example: Single-choice.} 
    \label{tab:my_label1}
\end{table*}

\begin{table*}[!t]
    \centering
    \begin{tabular}{|p{18cm}|}
\hline
          Use the following pieces of context to extract the information at the end. If you can't find the answer, just say that you don't know, don't try to make up an answer.\\   
\{summaries\}  \\
\\  
You can only answer from any of these values:\\

\{answer\_options\}  \\
\\
 Question: \{question\}  \\
\\
You can pick many options from the provided options, but you can only use each once. Return the final values when answering.\\

Answer:\\

\hline 

    \end{tabular}
    \caption{Prompt example: Multiple-choice.} 
    \label{tab:my_label2}
\end{table*}

\begin{table*}[!t]
    \centering
    \begin{tabular}{|p{18cm}|}
\hline
          Use the following pieces of context to extract the information at the end. If you can't find the answer, just say that you don't know, don't try to make up an answer.\\   
\{summaries\}  \\
\\
 Question: \{question\}  \\
\\
The answer can only be a number (or decimals) value between 0 to 1.  Just return the final value when answering.\\

Answer:\\

\hline 

    \end{tabular}
    \caption{Prompt example: Number-extraction.} 
    \label{tab:my_label3}
\end{table*}

\section{Evaluation and Results}

In this section, we discuss the study details, results and the analysis. 

\subsection{Evaluation details}
We performed empirical analysis on 13 studies covering 337 questions from the Cogtale dataset, and compare the generated answers using the LLM with the actual answer.
Below, we discuss the study selection criteria from the CogTale dataset, reasons for selecting the models and the metrics to evaluate the generated answers.

\textbf{Study Selection.} The CogTale dataset comprises studies categorized as verified and unverified, with verified studies encompassing published research papers. Among these, 52 studies were identified as verified. During manual scrutiny, we excluded studies where the correct answer was not provided among the options. Subsequently, to enhance the internal validity of our analysis and streamline the complexity, studies with more than one intervention or control component were excluded. Following these filtering steps, 40 studies remained. These were further divided into three sets, each containing 13, 13, and 14 studies. This subdivision aimed to analyze the sets separately and comprehend the model's performance on each. For this particular study, we focused on one of the set comprising 13 studies, reserving the others for future investigations. The titles of these selected studies are given in Table \ref{tab:studies} in Appendix A.


Among these 13 studies, most of the studies comprise 1 document, except one which consist of two documents. For each of the study, the dataset consist of 28 questions of various types requiring specific information. However, for a few studies that we used, some of the questions were not pertinent to the study and we do not use them for evaluation.  For instance, if a study is not of the randomized control type, specific questions lack a ground truth, and we omit asking these questions. Consequently, for each study, we pose only those questions for which the answer is present in the study. This approach resulted in a total of 337 questions evaluated across all the studies.

\textbf{Models.} We selected GPT-3.5-turbo and GPT-4 models to evaluate the document-based QA tasks on different question-types. These models are known for their natural language processing capabilities. GPT-3.5-turbo is known for its cost-effectiveness and efficient use of resources compared to some larger models, making it a practical choice for certain applications. On the other hand, GPT-4 represents a more recent and potentially more sophisticated iteration, offering an opportunity to assess the advancements in performance. We set the temperature variable of the models to 1 to ensure consistency in the analysis and results interpretation.

\begin{table*}[!t]
    \centering
    \begin{tabular}{|l|c|c|c|c|}
    \hline
     Category of Questions & N   &  Questions (\%) & GPT-3.5 & GPT-4 \\
     \hline
Yes-No & 143 &
42.43\% &
41.96\% & 
46.85\% \\
Single-choice & 73 &
21.66\% & 
38.36\% &
56.16\% \\
Single-choice (number) & 52 &
15.43\% & 
17.31\% & 
32.69\% \\
Multiple-choice & 43 &
12.76\% &  
4.65\% & 
25.58\% \\
Number-extraction & 26 &
7.72\% & 
26.92\% & 
19.23\% \\
\hline
    \end{tabular}
    \caption{Accuracy (in \%) of GPT-3.5-turbo and GPT-4 on CogTale dataset on the different type of questions. Here, Questions (\%) represents percentage of particular question-type. N represents the number of questions of the specific categories.}
    \label{tab:model_performance_cat}
\end{table*}

\textbf{Metric.} We conduct a comparison between the answers generated by the models and the actual ground truth answers, reporting whether they align. Given that the answers fall into either multiple-choice options or single numerical values, we perform a direct comparison between the generated answer and the ground truth, determining accuracy based on matching results. It is important to note that the ground truth results for the questions are provided within the dataset, offering a reliable reference for evaluation.


\begin{table*}[!t]
    \centering
    \begin{tabular}{|p{3.6cm}|p{1.5cm}|p{3.5cm}|p{2cm}|p{2.2cm}|p{2cm}|}
\hline


\textbf{Question} & \textbf{Category} & \textbf{Options} & \textbf{Actual answer} & \textbf{GPT 3.5} & \textbf{GPT-4}\\
\hline

Were the source(s) of participant recruitment adequately described?
&
Yes-No type
&
['Yes', 'No', 'N/A', 'Yes- Fully described', 'Yes- Partially described']
&
 Yes
&
 Yes- Partially described
&
 Yes- Partially described\\
\hline
 Does it mention any additional training given to participants about different strategies to use to perform and learn in the task?
&
Yes-No type
&
[Yes- single strategy, Yes- multiple strategies, No, N/A, Unclear, Yes]
&
 No
&
 Yes- multiple strategies
&
 Yes- multiple strategies\\

\hline

What is name of the intervention? & Single-choice & 
[Cognitive training  ,  Cognitive stimulation ,  Cognitive rehabilitation  ,  Brain training  ,  Brain games  ,  Cognitive exercise  ,  Mind training  ,  Memory training, Memory games, Memory rehabilitation, ..., Multi-Strategic Metamemory Training (MMT), Computerized Cognitive Training] &
Memory training & PM training & Cognitive training \\
\hline
What was the primary target of the intervention? &
Single-choice & 
[  Standardised cognitive task performance    ,   Standardised skill performance   ,   Everyday activity in context (e.g., banking)   ,   Psychoeducation, ...  Memory and learning  ,   Stress  ,...,   Not specified  ,   Speed/accuracy of processing  ] &
Memory and learning & The primary target of the intervention was the prospective memory (PM) component, specifically the enhancement of the prospective cue detection. & Visual imagery\\

\hline
 
What is the proportion of participants retained in the post-treatment experimental group? 
&
Number-extraction
&
-
&
0.88
&
0.88
&
0.76\\

\hline
What is the proportion of participants retained in the post-treatment experimental group? 
&
Number-extraction
&
-
&
0.80
&
0.75
&
0.80\\

\hline
What is the number of treatment/experimental conditions (not including control conditions)?
&
Single-choice (numbers)
&
[ 0 ,  1 ,  2 ,  3 ,  4 ,  5 ,  6 ,  7 ,  8 ,  9 ,  10 ,  11 ,  12 ,  13 ,  17 ,  16 ,  18 ,  27 ,  19 ,  14 ,  20 ,  23 ,  24 ,  21 ]
&
1
&
I don't know.
&
1
\\
\hline
What is the number of control conditions?
&
Single-choice (numbers)
&
[ 0 ,  1 ,  2 ,  3 ,  4 ,  5 ,  6 ,  7 ,  8 ,  9 ,  10 ,  11 ,  12 ,  13 ,  17 ,  16 ,  18 ,  27 ,  19 ,  14 ,  20 ,  23 ,  24 ,  21 ]
&
1
& 80
& 80 \\
\hline

    \end{tabular}
    \caption{Example of different question-type where the model generated incorrect answer. The questions shown here are randomly selected and are from different studies.} 
    \label{tab:example_yes_no1}
\end{table*}

\begin{table*}[!t]
    \centering
    \begin{tabular}{|p{2.6cm}|p{5.0cm}|p{2.5cm}|p{3cm}|p{3cm}|}
\hline

\textbf{Question:} & \textbf{Options} & \textbf{Actual answer} & \textbf{GPT 3.5} & \textbf{GPT-4}\\
\hline
\textit{What were the other targets of the intervention?}  &

 [ Standardised cognitive task performance   ,  Standardised skill performance  ,  Everyday activity in context (e.g., banking)  ,  Psychoeducation  ,  Goal-setting \& performance  ,  Pleasant activities  ,  Reminiscence  ,  Reality orientation  ,  Relaxation  ,  Conversation  ,  N/A  ,  well being ,  Mood ,  Physical Activity ,  Quality of Life (QoL) ,  Subjective Cognition (e.g., memory) ,  Activities of daily living ,  Stress management ,  Assertiveness ,  Use of memory aids ,  fMRI Imaging ,  Paper-and-pencil exercises ,  Behaviour ,  Global functioning ,  Social interaction ,  Self-confidence ,  Motivation ,  Psychiatric symptoms ,  Emotional symptoms ,  Strategy ,  Global cognitive function ,  Specific cognitive domains ,  Lifestyle ,  Memory and learning ,  Stress ,  PET Imaging ,  Subjective complaints ,  Apathy ,  Fatigue ,  Depression ,  Neuropsychiatric Behaviors  ,  Caregiver burden ,  Multisensory Stimulation ,  Not specified ,  Speed/accuracy of processing ]
&
Memory and learning
&
Goal-setting and performance ,  Psychoeducation ,  Pleasant activities, Standardised cognitive task performance ,  Standardised skill performance,  Conversation,  N/A ,  Reality  orientation ,  Everyday activity in context (e.g., banking),  Reminiscence ,  Relaxation ]

&
Specific cognitive domains, Global cognitive function,  Standardised skill  performance, 
Goal-setting and performance ,  Standardised cognitive task performance ,  Memory and learning, 
  Psychoeducation ,  Activities of daily living , Everyday activity in context (e.g., banking) \\

\hline 

What was the statistical tests used? & [ t-test ,  Analysis of variance  ,  Repeated Measures Analysis of Variance  ,  Linear Mixed Models  ,  Generalised Estimating Equations  ,  Regression ,  Mixed effects models ,  None  ,  N/A , "Fisher s Test",  Repeated Measures General Linear Model ,  Power calculation ,  Linear regression ,  Analysis of covariance ,  Mann-Whitney U-test ,  Pearson chi squared ,  Monte-Carlo significance ,  Jonckheere-Terpstra test , "Bonferroni s multiple comparison test",  Benjamini-Hochberg procedure - FDR ,  Kruskal-Wallis test ,  Wilcoxon test ,  Kolmogorov-Smirnov ,  ANCOVA ,  Mediating effect model ,  chi-square test , "Pearson s correlation",  Shapiro-Wilk test ,  constrained longitudional data analysis (cLDA) ] 
& t-test 
& Student’s t test for independent samples, Wilcoxon’s non-parametric test, Student’s t test for paired samples
& Kolmogorov-Smirnov, Student’s t test for independent samples, Wilcoxon test, Student’s t test for paired samples \\
\hline
    \end{tabular}
    \caption{Example of Multiple-choice from options question-type from a study, where both GPT 3.5 and GPT-4 selected more options as correct answer than the actual number of answer.} 
    \label{tab:example_mult_select}
\end{table*}

\subsection{Results and Analysis}
The overall accuracy of GPT-3.5-turbo and GPT-4 on these questions was found to be 31.45\% and 41.84\%, which shows GPT-4 outperformed GPT-3.5-turbo on the different question types in terms of correctly answering the questions. However, the overall accuracy of the two models are very low.

The performance of these models across the different question categories is shown in Table \ref{tab:model_performance_cat}. As shown, GPT-4 consistently outperformed GPT-3.5-turbo across various categories. Notably, both models exhibited superior performance in answering Yes-No and Single-choice type questions compared to other question types. However, their performance was less satisfactory for the Multiple-choice and Number Extraction question types. 

We look into some examples of yes-no type questions where the model failed to answer correctly. For example, in Table \ref{tab:example_yes_no1}, in the first question of yes-no type,  while the actual answer is \lq Yes', the models select \lq Yes-Partially described" as the answer. It is possible, the generated answer may have been correct when only yes and no options were present. However, in the second and third questions, it did not correctly answer the question.

For single-choice questions shown in the table, we found that GPT-3.5 selected options that were not present in the list of options. Similarly, GPT-4 also selected \lq Visual Imagery' option as the correct answer (for the fourth question) that was not present in the option list. These results could be due to models hallucinating answers. Additionally, when only one of the option is to be selected for the question \textit{What was the primary target of the intervention?}, GPT-3.5 instead answers the question in detail.

The examples of single-choice (numbers) and number-extraction questions are shown in the last 4 columns in Table \ref{tab:example_yes_no1}. The results show that the models have difficulty in answering numerical questions. Similarly, the single-choice (numbers) question types also had the same issue.

On multiple-type, GPT 3.5-turbo achieved very low accuracy (4.65\%) as compared to GPT 4 (25.58\%), demonstrating its poor performance on such questions. However, both the models achieved very low accuracy score as compared to single-choice categories. One of the reason for these results is that in some cases, both GPT-3.5 and GPT-4 selected more options than the actual number of correct answers, which causes this category to perform poorly. Table \ref{tab:example_mult_select} show examples of such cases. For this question, the options consist of multiple potential answers. Despite only one option being correct, both the models identified several options as correct. Additionally, it is important to highlight that the options selected by these models exhibited variations both in number and order.

\section{Discussion}
Our study on the CogTale dataset reveals that GPT-4 surpasses GPT-3.5-turbo in question-answering accuracy, achieving 41.84\% overall accuracy compared to 31.45\%. Both models excel in Yes-No and Single-choice questions but struggle with Multiple-choice and Number Extraction types. GPT-4, while an improvement, faces challenges in nuanced understanding and occasionally provides incorrect answers in Yes-No questions. For some single-choice questions, these models were observed to choose options that were not part of the provided set of options,  resulting in inaccuracies, possibly due to hallucination. Multiple-choice questions pose difficulties, with both models extracting more answers than necessary. Additionally, GPT-3.5-turbo's low accuracy (4.65\%) in Multiple-choice questions highlights limitations. Additionally, numerical understanding poses a distinct hurdle, reflected in the difficulty these models encountered in accurately addressing numerical type questions.

While the results reflect that the future models should address these issues by emphasizing nuanced comprehension and context awareness, it is also important to highlight potential challenges with the retriever component, which can impact the accuracy of information retrieval, and should be considered for comprehensive model improvement.

Furthermore, to run the same queries on these models, GPT-4 charged 26.54\$ which is 15 times higher than that of GPT-3.5-turbo (1.77\$). Therefore, while GPT-4 exhibits superior performance, the financial burden associated with its usage must be carefully weighed against the incremental gains in accuracy, especially in scenarios where cost-effectiveness is a paramount consideration. This economic dimension adds a layer of complexity to the decision-making process when selecting a model for practical applications.

\begin{table*}[!t]
    \centering
    \begin{tabular}{|c|p{16cm}|}
    \hline
      \textbf{No.}  &  \textbf{Title of the Studies selected from the Cogtale platform} \\
      \hline
      1 & Benefits of Training Working Memory in Amnestic Mild Cognitive Impairment: Specific and Transfer Effects -  \cite{carretti2013benefits}   \\
      \hline
      2 & Cognitive training in older adults with Mild Cognitive Impairment: Impact on cognitive and functional performance - \cite{brum2009cognitive} \\
      \hline
      3 & Effectiveness of a Visual Imagery Training Program to Improve Prospective Memory in Older Adults with and without Mild Cognitive Impairment: A Randomized Controlled Study - \cite{lajeunesse2022effectiveness}  \\
      \hline
      4 & Toward rational use of cognitive training in those with mild cognitive impairment - \cite{hampstead2023toward} \\
      \hline
      5 & Impact of metacognition and motivation on the efficacy of strategic memory training in older adults: Analysis of specific, transfer and maintenance effects - \cite{carretti2011impact} \\
      \hline
      6 & Repetition-lag training to improve recollection memory in older people with amnestic mild cognitive impairment. A randomized controlled trial - \cite{finn2015repetition} \\
      \hline
      7 & Effects of reality orientation therapy on elderly patients in the community - \cite{baldelli1993effects} \\
      \hline
      8 & Efficacy of a cognitive intervention program in patients with mild cognitive impairment - \cite{rojas2013efficacy} \\
      \hline
      9 & Computerized Structured Cognitive Training in Patients Affected by Early-Stage Alzheimer’s Disease is Feasible and Effective: A Randomized Controlled Study - \cite{cavallo2016computerized} \\
      \hline
      10 & Efficacy of the Ubiquitous Spaced Retrievalbased Memory Advancement and Rehabilitation Training (USMART) program among patients with mild cognitive impairment: a randomized controlled crossover trial - \cite{han2017efficacy} \\
      \hline
      11 & Cognitive rehabilitation combined with drug treatment in Alzheimer’s disease patients: a pilot study - \cite{bottino2005cognitive} \\
      \hline
      12 & Cognitive rehabilitation in patients with mild cognitive impairment - \cite{kurz2009cognitive} \\ 
      
      \hline
      
      13 & The PACE Study: A Randomized Clinical Trial of Cognitive Activity Strategy Training for Older People with Mild Cognitive Impairment - \cite{vidovich2015pace} \\
      \hline
      
    \end{tabular}
    \caption{Selected Studies for the experiments from the Cogtale platform}
    \label{tab:studies}
\end{table*}

\section{Conclusion and Future work}

In conclusion, this paper's evaluation of language models in question-answering tasks, facilitated by a dataset encompassing trial information and diverse inquiries and different question formats, has provided a nuanced perspective on their performance on information retrieval-based QA task. GPT-4 exhibited notable proficiency in certain question categories, adeptly grasping contextual cues to deliver coherent responses. However, the study also unveiled vulnerabilities when answering questions with multiple choices and  number extraction.
As language models continue to evolve, this evaluation serves as a guiding compass, navigating us toward more refined and versatile models that transcend the boundaries of textual and numerical comprehension, ultimately advancing the landscape of natural language processing. Notably, GPT-4's increased cost (26.54\$, 15 times higher than GPT-3.5-turbo's 1.77\$) prompts careful consideration of financial implications against the incremental gains in accuracy. 

For future work, we plan to evaluate this dataset with advanced prompting strategies such as CoT by \cite{wei2022chain} or others (\cite{singh2023tree}) . As some questions may require inferencing rather than directly extracting, different prompting strategies may help to retrieve the correct answer, and thereby improve the performance.  We also plan to study the affect of  retrievers on our study. Finally, it would be interesting to observe how models such as LLAMA 2 (\cite{touvron2023llama}), Gemini (\cite{team2023gemini}) and others perform on similar tasks. 


\section{Threat to Validity}
As we delve towards bridging the gap in evaluating LLMs on information retrieval QA task across diverse question formats using the CogTale dataset, it is imperative to acknowledge potential threats to the validity of our study's outcomes. The language models exhibit dynamic behavior, and a model update has the potential to enhance or diminish its performance on the dataset. Thus, the results on CogTale dataset may improve on a newer version of the GPT-4. 


In addition, we utilised a direct prompting strategy for the QA task. It would be worthwhile to investigate whether using different prompting techniques can significantly impact the results. Finally, our experiments were performed in the zero-shot setting. Few-shot setting may help to improve the performance of the LLM.

\appendix

\section{Studies Used}
The titles of the studies used in the analysis are presented in Table \ref{tab:studies}.


\bibliographystyle{elsarticle-harv} 
\bibliography{reference}






\end{document}